\documentclass[twocolumn, showpacs,preprintnumbers,amsmath,amssymb]{revtex4}
\usepackage{graphicx}
\usepackage{dcolumn}
\usepackage{bm}

\newcommand{\be}{\begin{equation}}
\newcommand{\ee}{\end{equation}}

\begin{document}

\title{Strain-stiffening in random packings of entangled granular chains}
\author{Eric Brown$^{1,2}$}
\author{Alice Nasto$^1$}
\author{Athanasios G. Athanassiadis$^1$}
\author{Heinrich M. Jaeger$^1$}
\affiliation{
$^1$James Franck Institute and Department of Physics, The University of Chicago, Chicago, IL 60637\\
$^2$School of Natural Sciences, University of California, Merced, CA 95343
}
\date{\today}

\begin{abstract} 

Random packings of granular chains are presented as a model polymer system to investigate the contribution of entanglements to strain-stiffening in the absence of Brownian motion.  The chain packings are sheared in triaxial compression experiments.  For short chain lengths, these packings yield when the shear stress exceeds a the scale of the confining pressure, similar to packings of spherical particles.  In contrast, packings of chains which are long enough to form loops exhibit strain-stiffening, in which the effective stiffness of the material increases with strain, similar to many polymer materials.  The latter packings can sustain stresses orders-of-magnitude greater than the confining pressure, and do not yield until the chain links break.  X-ray tomography measurements reveal that the strain-stiffening packings contain system-spanning clusters of entangled chains. 

\end{abstract}

\pacs{83.80.Fg, 81.70.Tx, 62.20.mm, 61.41.+e}

\maketitle

Most materials, ranging from crystalline metals to piles of granular material, become weaker the further they are strained.  On the other hand, many polymeric materials are known to strain-stiffen, in which the effective stiffness of the material increases as the material is strained further \cite{Treolar49}.  Theories suggest strain-stiffening could depend on many factors including  chain stiffness, density, temperature, strain rate, and in particular on structures such as entanglements between different chains \cite{BJ95, HR06, VLM09, HO10}.  However, it has not been possible to directly measure entanglements in experiments because in polymers these structures occur on very small scales.  In this letter we present a new experimental approach to investigate the role of entanglements in strain-stiffening using a model system of granular chains consisting of millimeter-scale beads connected by flexible links.  Such a macroscopic system has advantages over molecular polymer systems for investigating entanglement.  First, the macroscopic size allows for imaging to measure the precise positions of each particle and link in the structure.  Second, we can isolate entanglement effects from temperature and strain rate dependent effects because the macroscopic chains have no inherent time scales due to Brownian motion or relaxation.

Granular chains have been considered as model polymers in previous studies to characterize the packing structure around the jamming transition for chain packings \cite{KFL09, ZCRJN09, LRR11}. Such chains form tight loops when packed which defines a characteristic loop size and persistence length in analogy to polymers. The free volume in the packing was found to increase with and level off when the chain length exceeded the persistence length, analogous to the behavior of the glass transition temperature for polymers  \cite{ZCRJN09}.  Similarly, tumbling strings have been considered as a macroscopic model polymer system, in which knotting was found to exhibit statistics analogous to thermodynamic systems \cite{RS07}.  While these works characterized the structures of macroscopic model polymer systems, the stress response of such systems have not previously been characterized.  In this letter, we report stress measurements to demonstrate a more direct connection with polymer materials; specifically that they exhibit similar strain-stiffening.  We use x-ray tomography to measure the precise packing structure and identify entanglements to demonstrate a quantitative connections between entanglements and strain-stiffening. 


\begin{figure}
\includegraphics[width=2.4in]{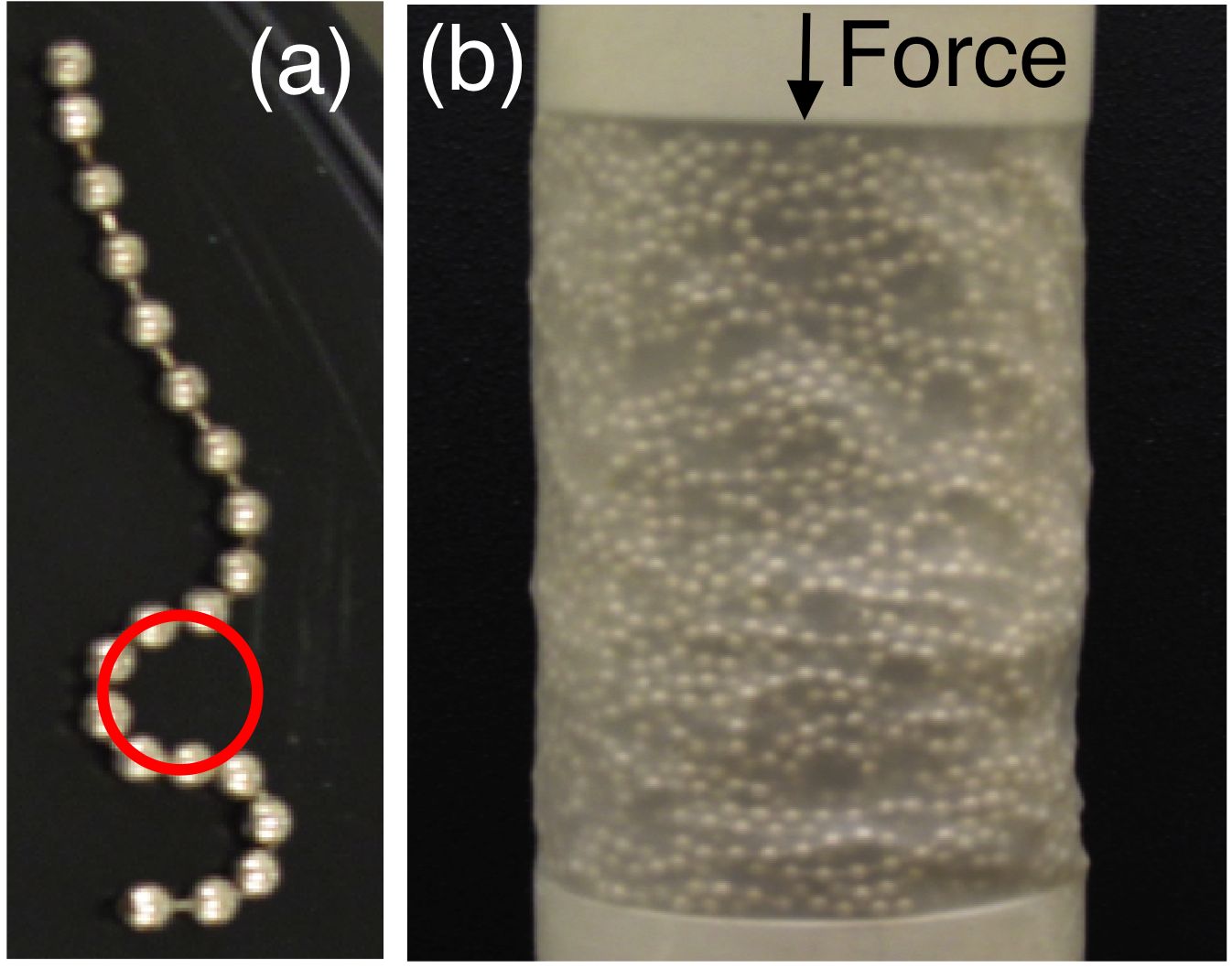}
\caption{(a)  An individual chain with length $N=20$ beads.  The minimum loop size of $\xi=8$ beads is highlighted by the red circle. (b) Cylindrical packing of chains with $N\gg \xi=8$ inside a thin latex membrane used for triaxial compression measurements.  Many tight loops near the minimum loop size can be seen in the packing through the membrane.
}
\label{fig:chains}
\end{figure}

As a model material, we used macroscopic granular chains consisting of hollow spherical brass beads shown in Fig.~\ref{fig:chains}a.  The beads are flexibly connected into chains by enclosing dog-bone-shaped brass links.  These connections have essentially zero stiffness for small bend angles, but they have a maximum bend angle beyond which the chains will not bend without plastic deformation.  This defines a minimum loop circumference $\xi$ the chains can bend into, as shown in Fig.~\ref{fig:chains}a.  We used two sets of chains with beads of slightly different diameters of 1.9 mm and 2.1 mm which causes them to have different minimum loop circumferences of $\xi = 8$ and $\xi=11$ beads, respectively.  In contrast to polymer chains, these chains have a non-linear response to bending as described above, and they are far too large to experience Brownian motion.

We measured the mechanical response of granular chain packings using triaxial compression \cite{LW69}.  The grains were poured into a flexible elastic membrane in a cylindrical shape with solid end caps on the top and bottom, as shown in Fig.~\ref{fig:chains}b.  An Instron materials tester was used to quasi-statically push the top end cap downward while measuring its height and the required force for samples with diameter and initial height of 50 mm.  This average compressive stress $\tau$ is calculated as the force over the cross-sectional area of an end cap and the compressive strain $\gamma$ is calculated as the height change (positive downward) over the initial height. 


\begin{figure}
\includegraphics[width=3.2in]{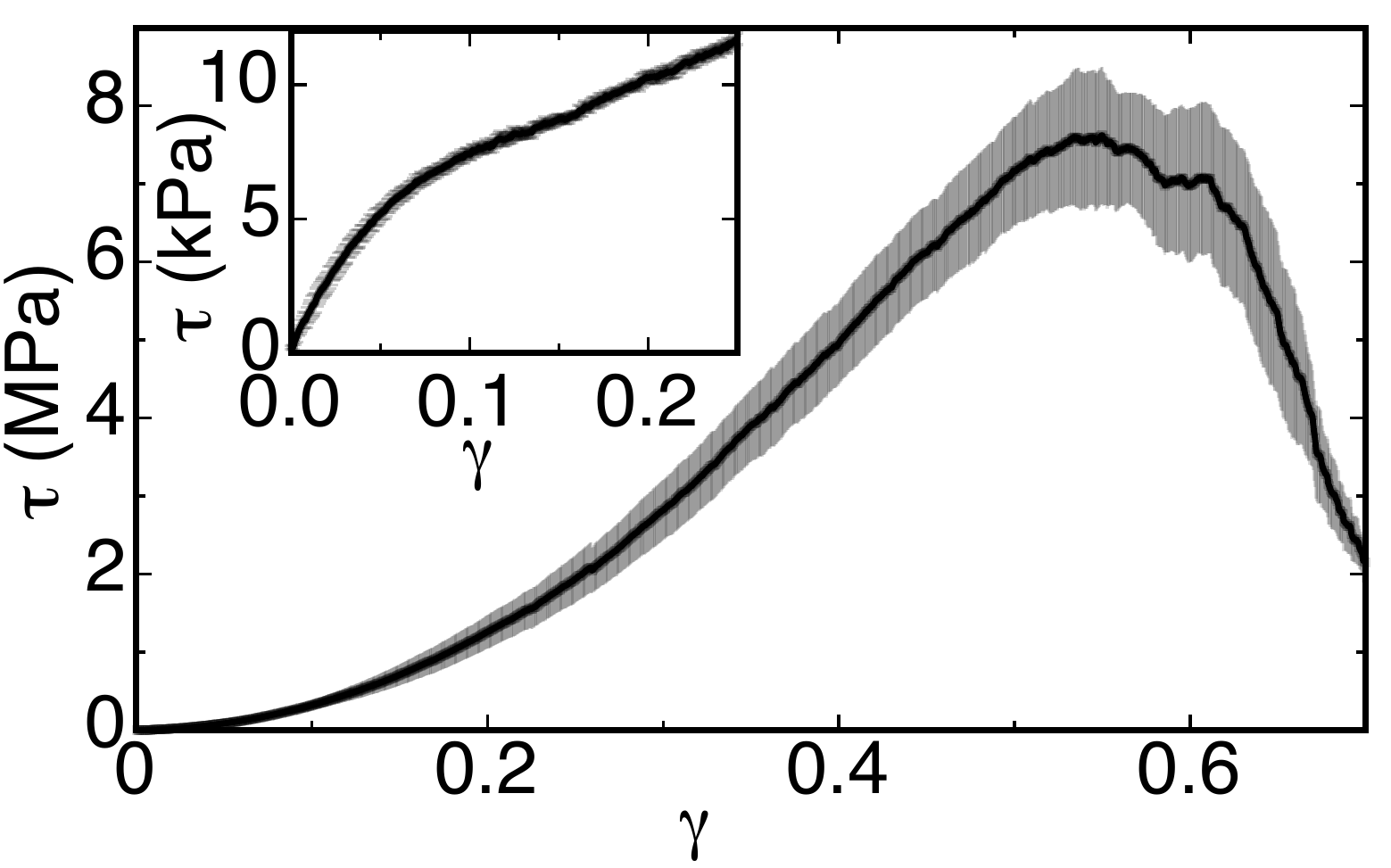}
\caption{Stress $\tau$ vs. strain $\gamma$ for packings of chains with length $N\approx10^4$.  The shaded bands represent 1 standard deviation based on 5 repeated measurements. Strong strain-stiffening is seen as the positive curvature in $\tau(\gamma)$.  Inset:  $\tau(\gamma)$ for $N=1$ (unlinked beads).
}
\label{fig:longchain}
\end{figure}

We show in Fig.~\ref{fig:longchain} the stress-strain relation $\tau(\gamma)$ for a packing of chains with length $N\approx10^4$ beads,  much longer than their minimum loop circumference $\xi=8$.  Enormous strain-stiffening can be seen as the region of positive curvature of $\tau(\gamma)$.  A striking feature is that the stress reaches the order of 10 MPa before the packing fails, which is indicated by the flattening of the stress-strain relation where the packing is able able to shear without supporting additional stress.  This ultimate strength is on the order of 1/10 the strength per density of the solid brass that makes up the chains (this packing is about 20\% brass by volume) and comparable to polymer rubbers.  Such high strength per density in a pourable, shapable granular material could have unique engineering applications.  In contrast, unconnected granular materials typically fail when the shear stress exceeds the confining pressure at the boundary in the absence of interparticle attractions \cite{LW69}.  In our experiments, the total confining stress coming from gravity and the elastic membrane is only about 10 kPa, which is the maximum stress scale reached for $N=1$ (unlinked beads) in the inset of Fig.~\ref{fig:longchain} (the positive slope for $\gamma \stackrel{>}{_\sim} 0.05$ matches the contribution of the membrane stiffness as it is deformed).  The chain packings exceed this stress by a factor of $\sim 10^3$.  The strength of polymer materials is partly attributed to the extensional or bending strength of individual chains.  Indeed, during the compression of the long chains in Fig.~\ref{fig:longchain} we started to hear chains break at the point where the maximum stress was reached, and after each measurement was done we counted approximately 10\% of the links in the chains to be broken.   This breaking at such stresses suggests that -- similar to polymer systems -- the material strength of the chains themselves contributes and the packing can only shear and fail by breaking links in the chains, which would require the packing structure to produce some self-confinement.

\begin{figure}
\includegraphics[width=3.in]{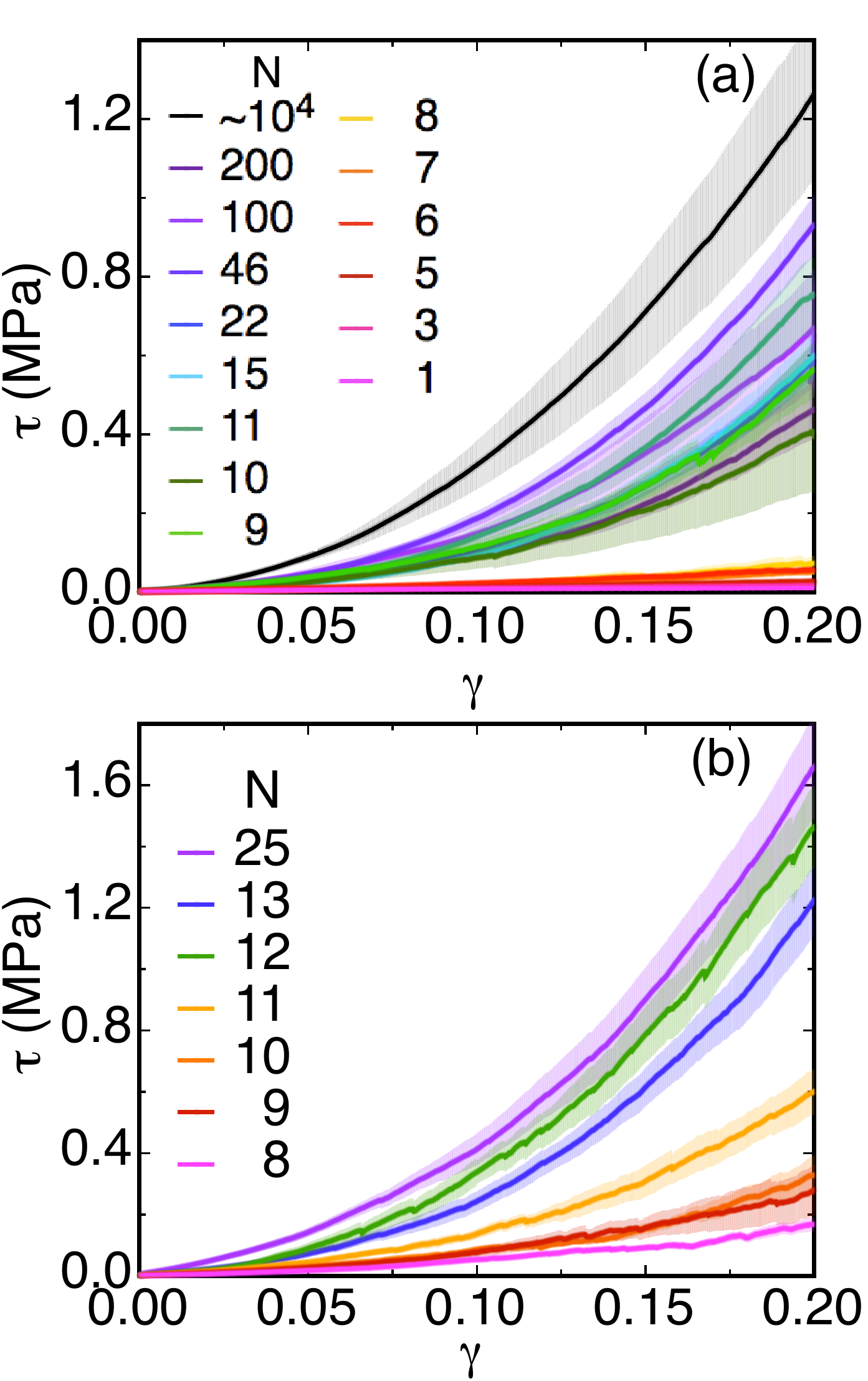}
\caption{Stress $\tau$ vs. strain $\gamma$ for different chain lengths $N$ shown in the key. (a)  For chains of minimum loop circumference $\xi=8$. (b) For $\xi=11$.  In both panels, the data for $N > \xi$ (cool colors) group into a band which exhibits strong strain-stiffening.  
}
\label{fig:stressstrain}
\end{figure}

To investigate the transition from granular behavior in the limit of unlinked beads with $N=1$ and strain-stiffening in the limit of large $N$, we show stress-strain curves for packings of chains with different lengths $N$ in Fig.~\ref{fig:stressstrain}.  Chains of minimum loop circumferences $\xi=8$ and $\xi=11$ are shown in panels a and b, respectively.  It is seen in both panels that the data fall into two distinct bands:  the upper band of data with chain lengths $N > \xi$ exhibits strong strain-stiffening, while the lower band of data with $N\le \xi$ exhibits typical granular strain-softening or much weaker strain-stiffening.  The fact that the transition length between these two bands scales with and is about equal to the minimum loop size $\xi$ in both cases suggests a relationship between strain-stiffening and whether the chains form loops in the packings.

\begin{figure}
\centerline{\includegraphics[width=3.in]{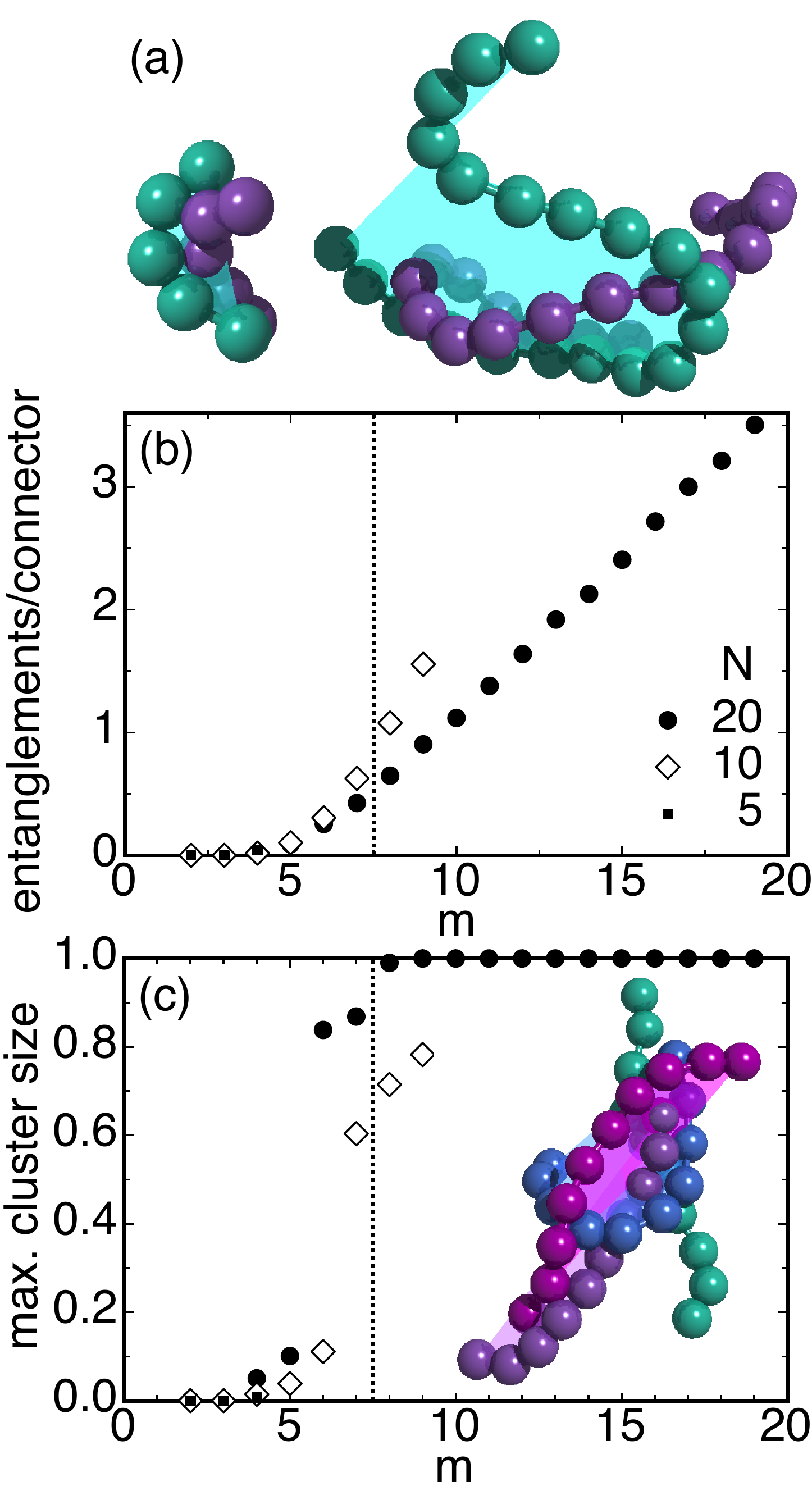}}
\caption{(a) Examples of entanglements for chain lengths $N=5$ (left) and $N=20$ (right).  The green chains are entangling the purple chains.  The shaded region indicates the entanglement manifold connecting the ends of the chains ($m=N-1$). (b) Average number of entanglements per connector as a function of connector separation $m$ for $\xi=8$.  The key indicates the chain lengths $N$.  (c) Maximum size of entangled clusters as a fraction of the total number of chains in the sample.  The dotted line indicates the onset of strong strain-stiffening, which is only found to occur when the connector separation is large enough to that there are entangled clusters consisting of the majority of chains in the system.  The inset image shows an example of a cluster of 4 chains with 5 entanglements between them for $N=10$.}
\label{fig:entanglements}
\end{figure}

To investigate the role of entanglement in this relationship, we used x-ray tomography to precisely reconstruct the positions of each bead and link in three dimensions.  For these experiments, we used aluminum chains with bead diameter 2.5 mm and $\xi = 7.5$ in cylindrical packings with diameter and height of 35 mm each.  We show results for chain lengths $N=5$, $10$ and $20$ each compressed to a strain of $\gamma = 0.2$.

We define an entanglement to occur if one chain wraps partially around another.  Motivated by string knotting analysis \cite{RS07}, we first draw an imaginary line called the connector between any two beads of the entangling chain separated by a distance of $m$ beads along the contour of the chain.  We then define an entanglement manifold as the minimal two-dimensional manifold bounded by the connector and the links that connect the intervening beads.  If another chain crosses through this manifold, it is counted as entangled by the entangling chain.  Examples of entanglements along with the manifolds are shown in Fig.~\ref{fig:entanglements}a.  These particular examples would not be counted as entanglements based on typical polymer algorithms where entanglements are counted if chains catch on each other if they are contracted \cite{LKMFK09}.  In contrast with elastic polymers, our granular chains are highly constrained from significant rearrangements due to their high packing density, so we expect these examples are still able to produce significant constraints against shearing from entanglement.  Regardless, the majority of our entanglements would satisfy the typical requirements, and it is usually expected that different algorithms for measuring entanglement tend to produce qualitatively similar statistics \cite{LKMFK09}.  

\begin{table}
\begin{tabular}{|l|r|r|}
 \hline
minimum loop circumference $\xi$ & $8.0\pm 0.5$ & $11.0\pm 0.3$\\ \hline
min. length for strong strain-stiffening  & $8.0\pm 0.5$ & $11.0\pm 0.5$\\ \hline
min. length for majority clusters & $6.5\pm1.0$ & -- \\ \hline
\end{tabular}
\caption{Comparison of different length scales.  The chain length required for strong strain-stiffening is just above that required for most of the chains to be entangled in a single cluster, and consistent with the minimum loop circumference of the chains. 
}
\label{tab:lengths}
\end{table}

We count all of the entanglements for each possible connector between beads and plot the average number of entanglements per connector as a function of separation distance $m$ in Fig.~\ref{fig:entanglements}b \cite{connectornote}.  Regardless of chain length $N$, for connector separations of $m \le 3$, no entanglements are observed because the size of the beads restricts other chains from fitting inside even the largest possible entanglement manifolds.  For larger $m$, the average number of entanglements per connector increases monotonically with $m$.   Since chains that are longer compared to their minimum loop size can be bent further, the area of the entanglement manifold tends to similarly increase with $m$ and it is more likely in a random packing that other chains will cross this manifold to become entangled.  Thus, the entanglement manifold method provides an intuitive way to understand why the probability of entanglements increases with separation or chain length.

The increase in the number of entanglements per connector with separation distance seen in Fig.~\ref{fig:entanglements}b could help explain why only the longer chains strain-stiffen.  The longer chains with $N=10$ and 20 which strain-stiffen reach more than 1 entanglement per connector for larger connector separations, while the shorter chains with $N=5$ which do not strain-stiffen have no more than 0.04 entanglements per connector. Since the probability of entanglements as a function of separation distance is similar for different $N$, it seems likely that the reason that the $N=5$ chains do not entangle more is simply that they are too short compared to the minimum loop size $\xi=8$ for them to bend much and thus have entanglement manifolds of large enough size to result in many entanglements.

If entanglements are responsible for constraints that prevent failure by shear, then at minimum the entanglements would have to connect into system-spanning clusters to avoid weak points in the packing that could still shear.  We identify a cluster as subset of chains in entanglements that are transitively connected.  An example of a cluster of four entangled chains is shown in the inset of Fig.~\ref{fig:entanglements}c. The largest cluster size found for each $N$ and $m$ is plotted in Fig.~\ref{fig:entanglements}c.  For the strain-stiffening chains with $N=10$ and 20, the largest cluster size jumps sharply to over 50\% of the chains in the packing at $m=6$ and 7, respectively.  This is just below the chain length $N=8$ required for strong strain-stiffening (these length scales are summarized in Table \ref{tab:lengths}).  These cluster sizes also large enough that they can span the system in several directions.  This is necessary for there to be enough constraints from entanglements in those clusters to fully constrain the packing against shear.  In contrast, the $N=5$ chains which do not strain-stiffen also do not have large enough clusters of entanglements to span the system. 


To summarize, we demonstrated that random packings of granular chains can be used as a model system that exhibits strain-stiffening as in polymer materials, with the additional ability to precisely measure the structure of entanglements.  These chains exhibit strain-stiffening if they are longer than the minimum loop circumference they make in packings, which allows them to form system-spanning clusters of entanglements.  Since the strength of granular materials is usually limited by the scale of the confining pressure at the boundaries, we propose that for the chain packings to reach the orders-of-magnitude higher strength, the confinement at the boundary must be functionally replaced by a self-confinement due to the system-spanning clusters of entanglements.  Strain-stiffening then results if these entanglements tighten up under strain, eventually locking up to prevent further shear unless the links break.  This picture contrasts with recent theoretical arguments that strain-stiffening in polymers can be explained without entanglement.   However, those arguments only explained much weaker strain-stiffening in which the stress increase is less than an order of magnitude \cite{VLM09, HO10}, so no additional confining stress mechanism was necessarily in those cases.  The questions of whether entanglement is the only way to get strong strain-stiffening, and how to quantitatively predict strain-stiffening from entangled structures, remain open for future work.

\section{Acknowledgements}

We thank Peter Eng and Mark Rivers for their assistance with x-ray tomography, Ling-Nan Zhou for his particle tracking code, and Nicholas Rodenberg and Dylan Murphy for preliminary measurements.  We acknowledge the Advanced Photon Source general user program at Argonne National Laboratory for  providing beam time.  This work was supported by the NSF MRSEC program under DMR-0820054.


\begin{references}
\bibitem{connectornote} For each connector separation length $m$, there are $N-m$ possible pairs of endpoints of the connector.  We count an entanglement each time one of these endpoint pairs results in an entanglement, but then normalize by the $N-m$ possible connectors to get the average number of entanglements per connector.  This normalization prevents biasing the probably of entanglement for different connector separations due to the different number of possible connectors.
\end{references}
\end{document}